\documentclass[10pt]{bmc_article}    

\usepackage{amssymb, amsmath, amsthm}   
\usepackage{cite} 
\usepackage{url}  
\usepackage{ifthen}  
\usepackage{multicol}   
\usepackage[utf8]{inputenc} 
\usepackage{graphicx}   
\usepackage[draft]{fixme}
\newcommand{\eat}[1]{}
\newcommand{\PP}{\mathbb{P}}
\newcommand{\Tref}{T_{\mathrm{ref}}}
\newcommand{\Lcal}{\mathcal{L}}
\newcommand{\Lbayes}{\Lcal_{\mathrm{Bayes}}}
\newcommand{\Lml}{\Lcal_{\mathrm{ML}}}
\newcommand{\pplacer}{\texttt{pplacer}}
\newcommand{\Pplacer}{\texttt{Pplacer}}
\newcommand{\placeviz}{\texttt{placeviz}}
\newcommand{\Placeviz}{\texttt{Placeviz}}
\newcommand{\placeutil}{\texttt{placeutil}}
\newcommand{\Placeutil}{\texttt{Placeutil}}

\newcommand{\pro}{\emph{Prochlorococcus}}

\newcommand{\psbA}{\emph{psbA}}
\newcommand{\psbD}{\emph{psbD}}

\newcommand{\ncogs}{631}

\newcommand{\bmcSubmit}[1]{}
\newcommand{\bmcNoSubmit}[1]{}
\newcommand{\arxiv}[1]{#1}

\newcommand{\figapplicationEDPL}{1}
\newcommand{\INCapplicationEDPL}{\includegraphics[height=6in]{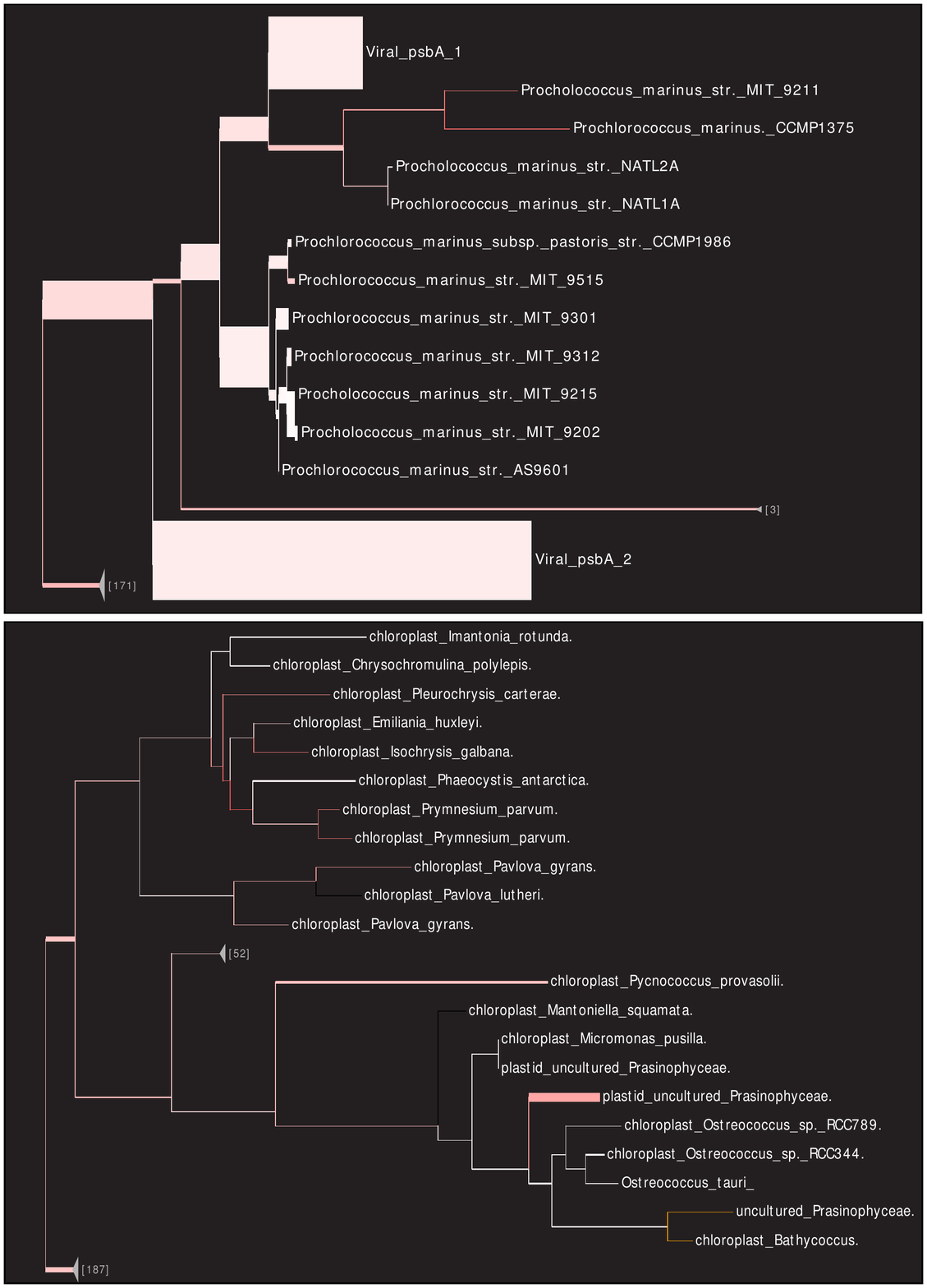}}
\newcommand{\LEGapplicationEDPL}{
\Pplacer\ example application using \psbA\ reference sequences and the corresponding recruited Global Ocean Sampling \cite{venter2004egs} (GOS) sequences showing both number of placements and their uncertainty.
Branch thickness is a linear function of the log-transformed number of placements on that edge, and branch color represents average uncertainty (more red implies more uncertain, with yellow denoting EDPL above a user-defined limit).
The upper panel shows the \pro\ clade of the tree.
The lower panel shows a portion of the tree with substantial uncertainty using the EDPL metric.
\Placeviz\ output viewed using Archaeopteryx \cite{archaeopteryx}.
}

\newcommand{\figspeed}{2}
\newcommand{\INCspeed}{\includegraphics[width=5.5in]{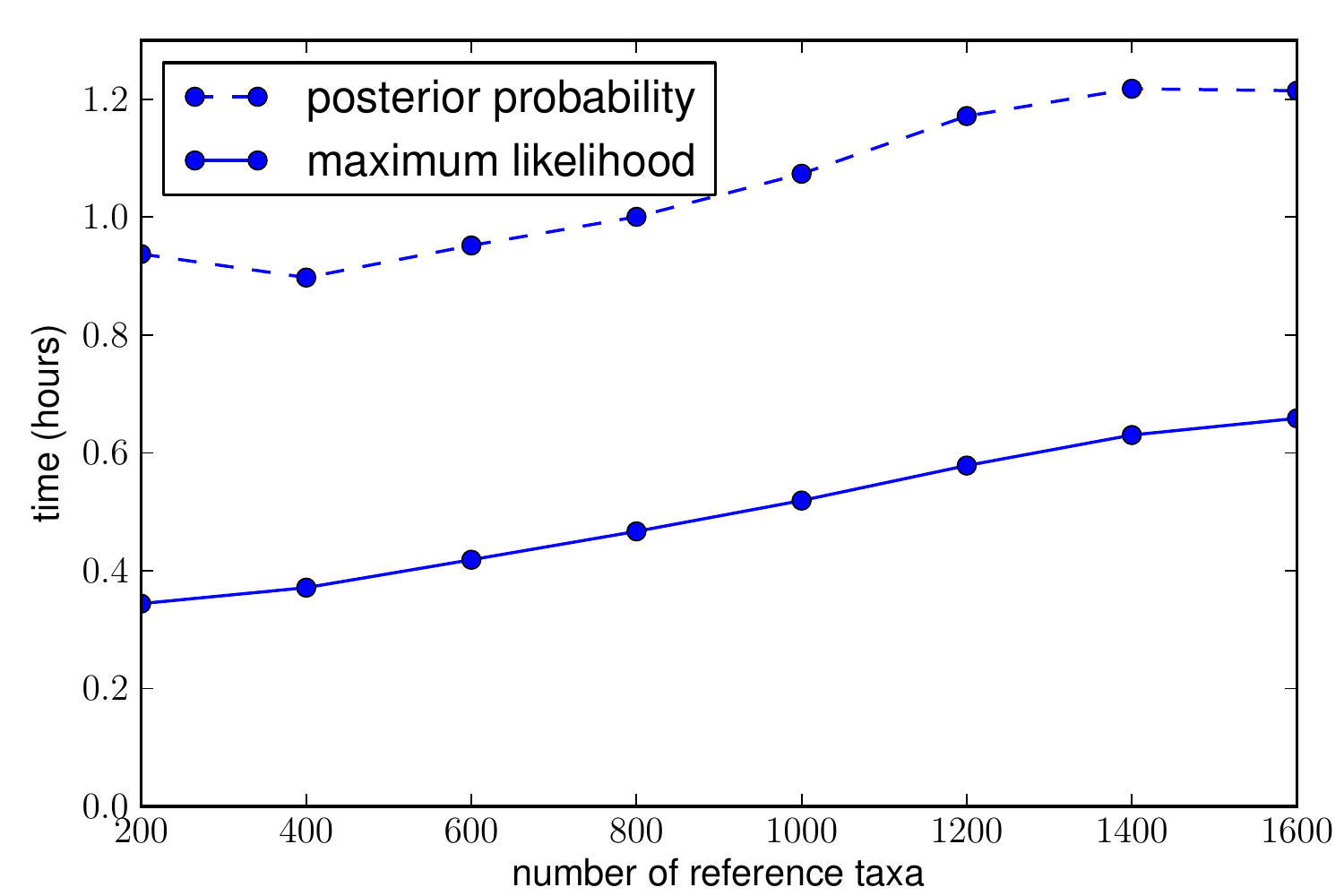}}
\newcommand{\LEGspeed}{
Time to place 10,000 16s rRNA reads of median length 198nt onto a reference phylogenetic tree, with a 1287 nt reference alignment. 
Tests run on an Intel Xeon @ 2.33 Ghz.
}

\newcommand{\figmem}{3}
\newcommand{\INCmem}{\includegraphics[width=5.5in]{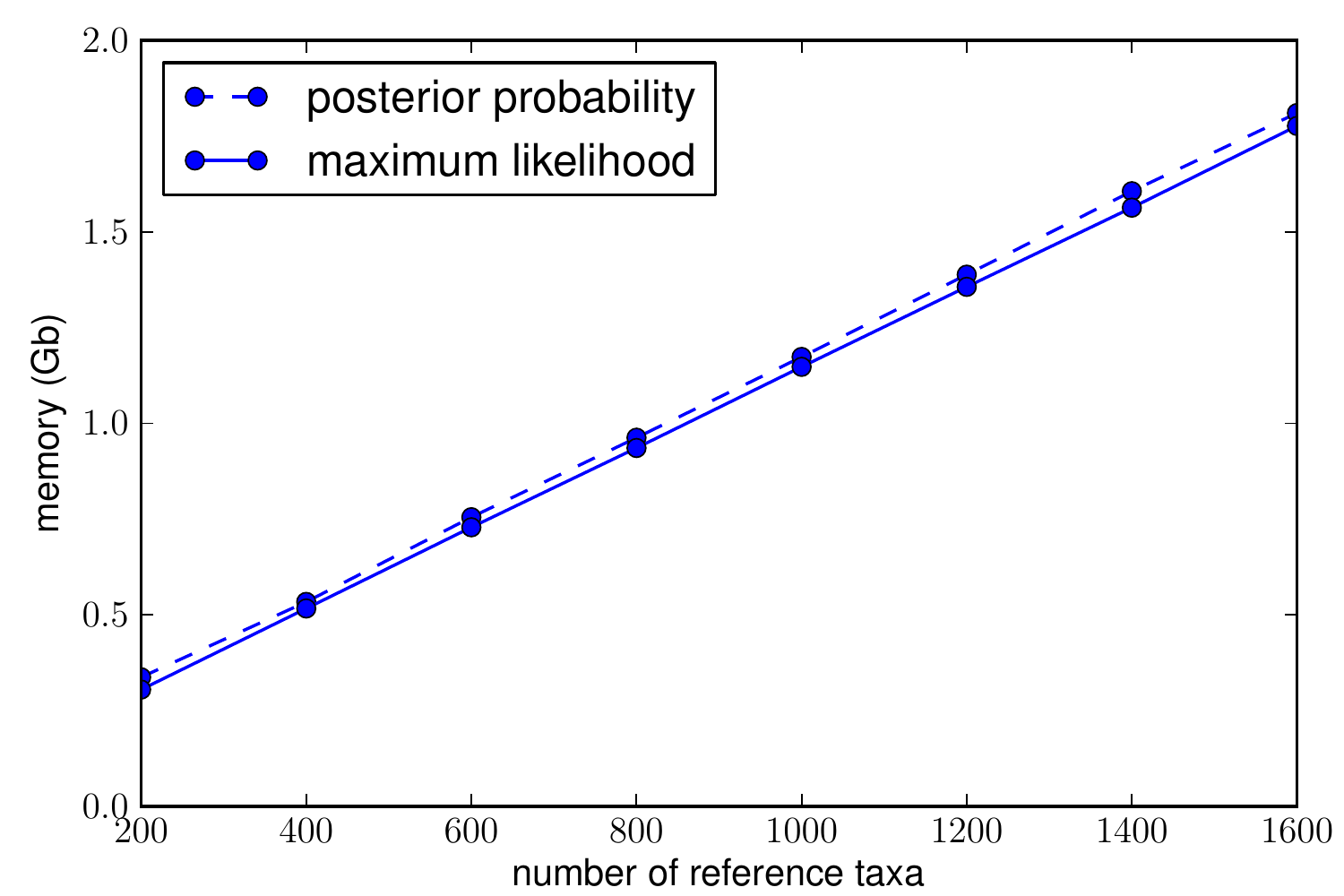}}
\newcommand{\LEGmem}{
Memory required to place 10,000 16s rRNA reads of median length 198nt onto a reference phylogenetic tree, with a 1287 nt reference alignment. 
Tests run on an Intel Xeon @ 2.33 Ghz.
}

\newcommand{\figEDPL}{4}
\newcommand{\INCEDPL}{\includegraphics[height=3in]{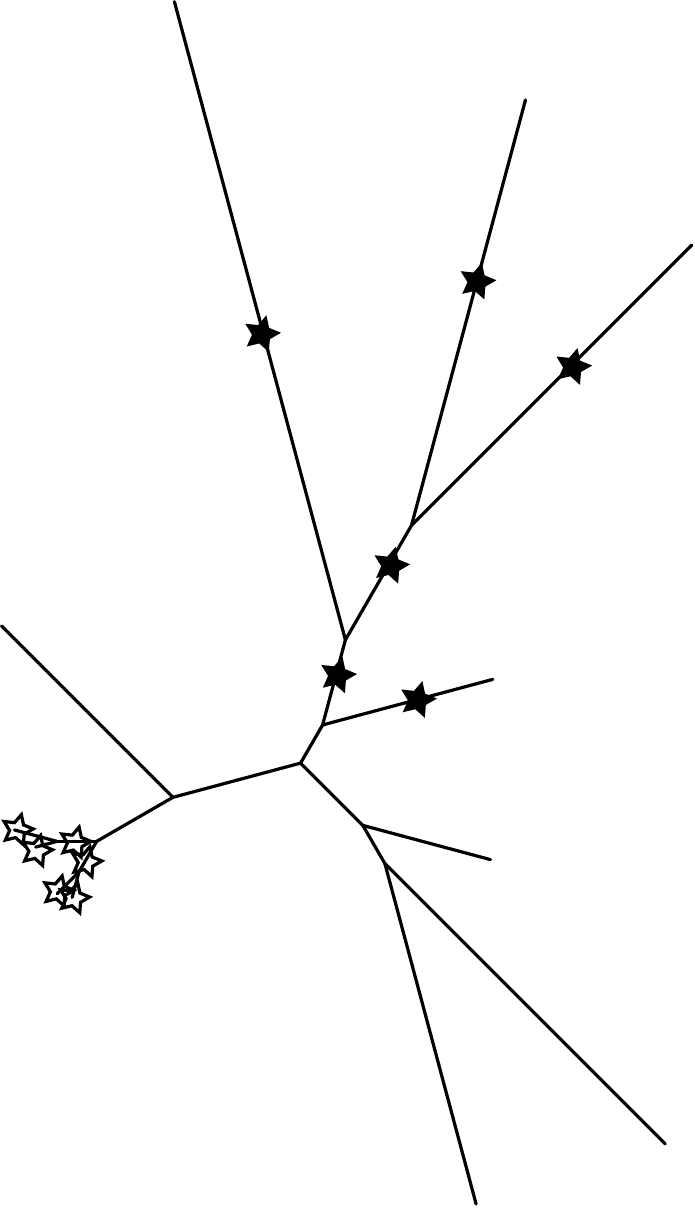}}
\newcommand{\LEGEDPL}{
The Expected Distance between Placement Locations (EDPL) uncertainty metric can indicate if placement uncertainty may pose a problem for downstream analysis. 
The hollow stars on the left side of the tree depict a case where there is considerable uncertainty as to the exact placement edge, but the collection of possible edges all sit in a small region of the tree.
This local uncertainty would have a low EDPL score.
The full stars on the right side of the diagram would have a large EDPL, as the different placements are spread widely across the tree.
Such a situation be flagged for special treatment or removal.
}

\newcommand{\figapplication}{5}
\newcommand{\INCapplication}{\includegraphics[height=6in]{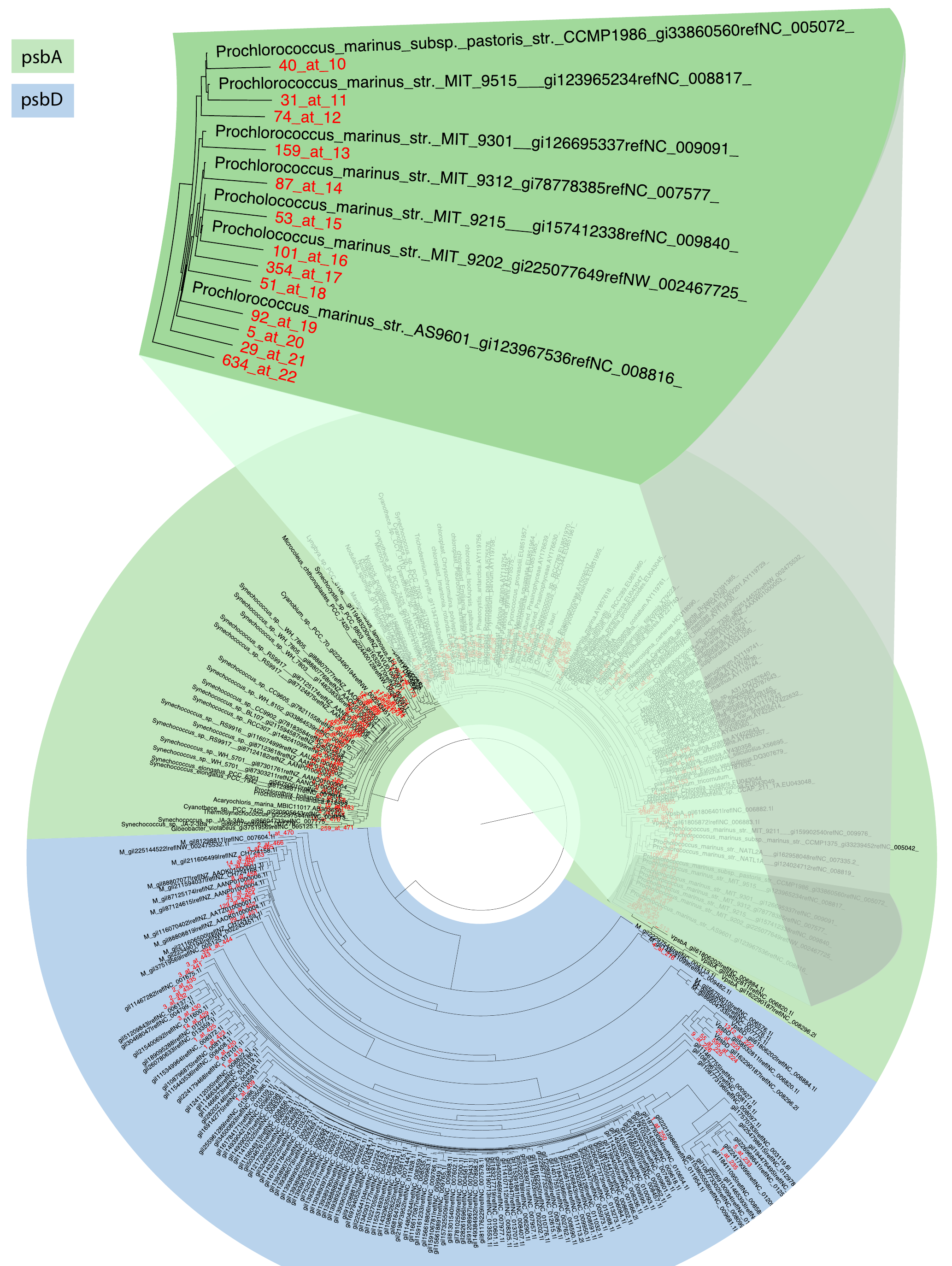}}
\newcommand{\LEGapplication}{
Placement visualization of same results as in Figure~\figapplicationEDPL.
The notation ``15\_at\_4'', for example, means that 15 sequences were placed at internal edge number 4.
These edge numbers can then be used to find the corresponding sequences in the \texttt{.loc.fasta} file.
\Placeviz\ output viewed using FigTree \cite{figtree}.
}

\newcommand{\figmaincog}{6}
\newcommand{\INCmaincog}{\includegraphics[width= 5.5in]{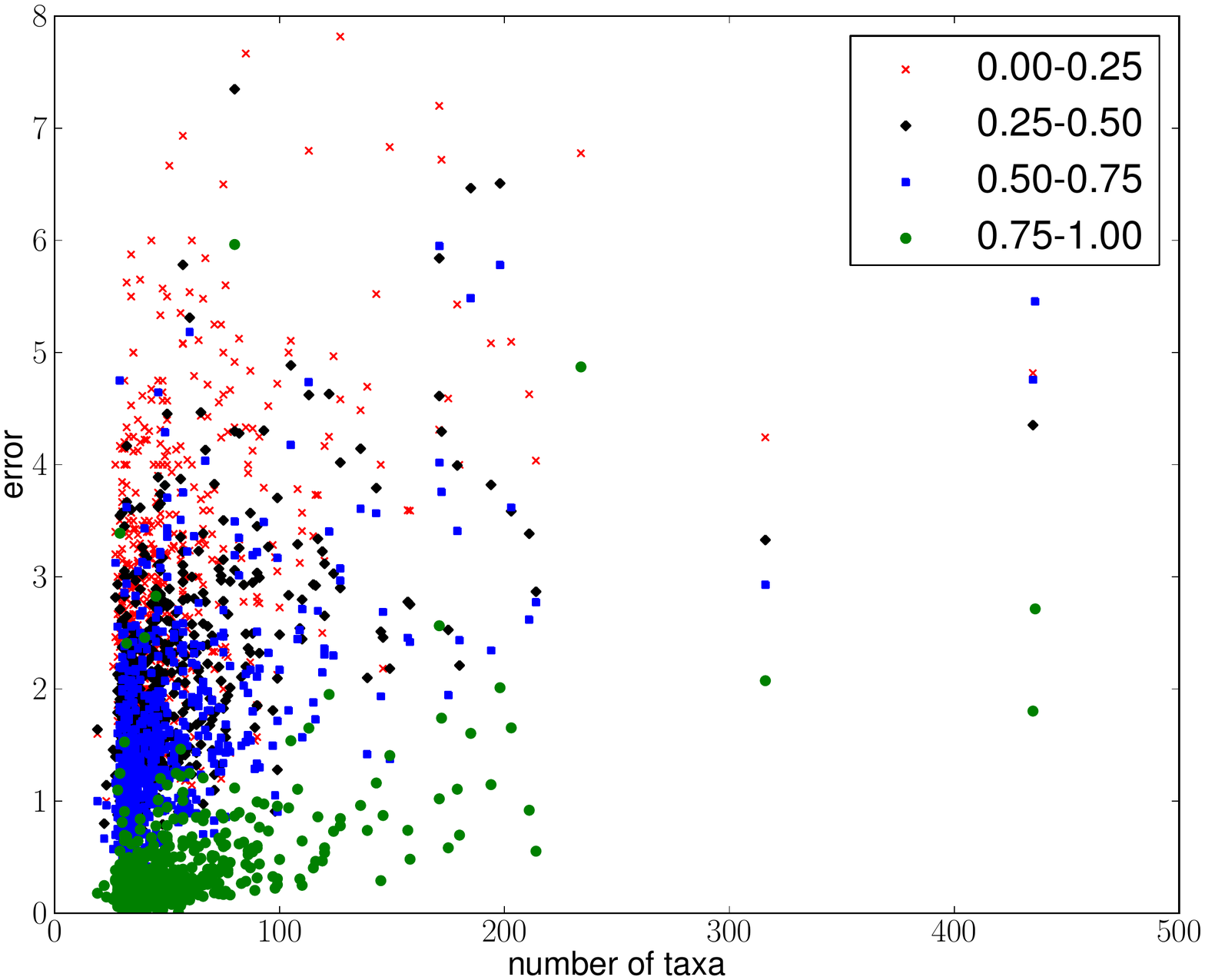}}
\newcommand{\LEGmaincog}{
Error analysis from a simulation study using \ncogs\ COG alignments.
Ten reads were simulated from each taxon of each alignment, and then binned according to the likelihood weight ratio of their best placement; ranges for the four bins are indicated in the legend.
There is one scatter point in the plot for each bin of each alignment: the $x$-axis for each plot shows the number of taxa in the tree used for the simulation, and the $y$ axis showing the average error for that bin.
For example, a point at $(100,1.2)$ labeled $0.5-0.75$ indicates that the set of all placements for an alignment of 100 taxa with confidence score between $0.5$ and $0.75$ has average error of $1.2$.
As described in the text, the error metric is the number of internal nodes between the correct edge and the node placement edge.
}

\newcommand{\figsistercog}{7}
\newcommand{\INCsistercog}{\includegraphics[width= 5.5in]{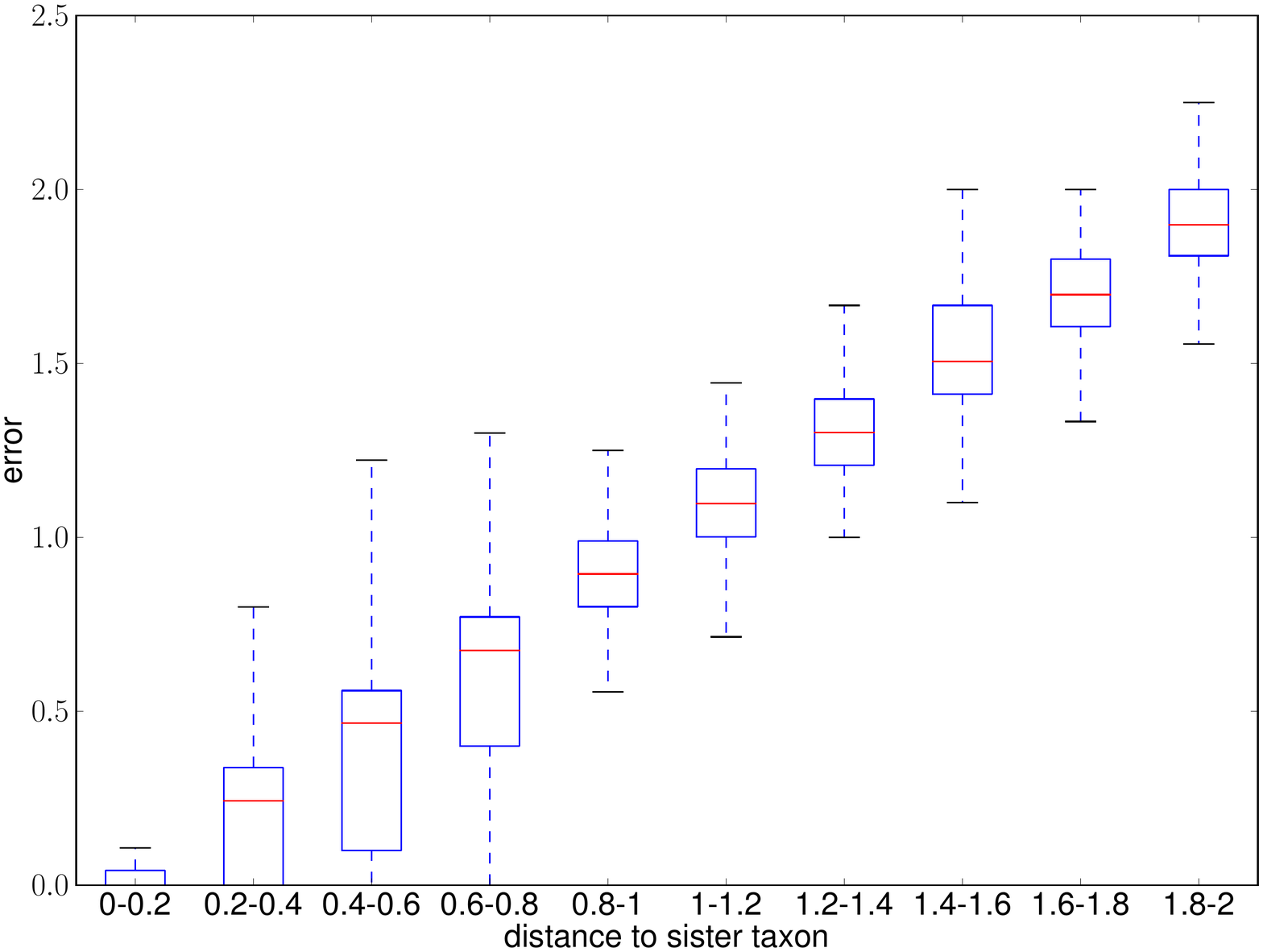}}
\newcommand{\LEGsistercog}{
The relationship between accuracy and phylogenetic distance to the sister taxon for the COG simulation.
For each taxon in each alignment, the phylogenetic distance to the closest sister taxon was calculated, along with the average placement error for the ten reads simulated from that taxon in that alignment.
The results were binned and shown in boxplot form, with the central line showing the median, the box showing the interquartile range, and the ``whiskers'' showing the extent of values which are with 1.5 times the interquartile range beyond the lower and upper quartiles.
Outliers eliminated for clarity.
}

\newcommand{\figEPA}{8}
\newcommand{\INCEPA}{\includegraphics[width= 5.5in]{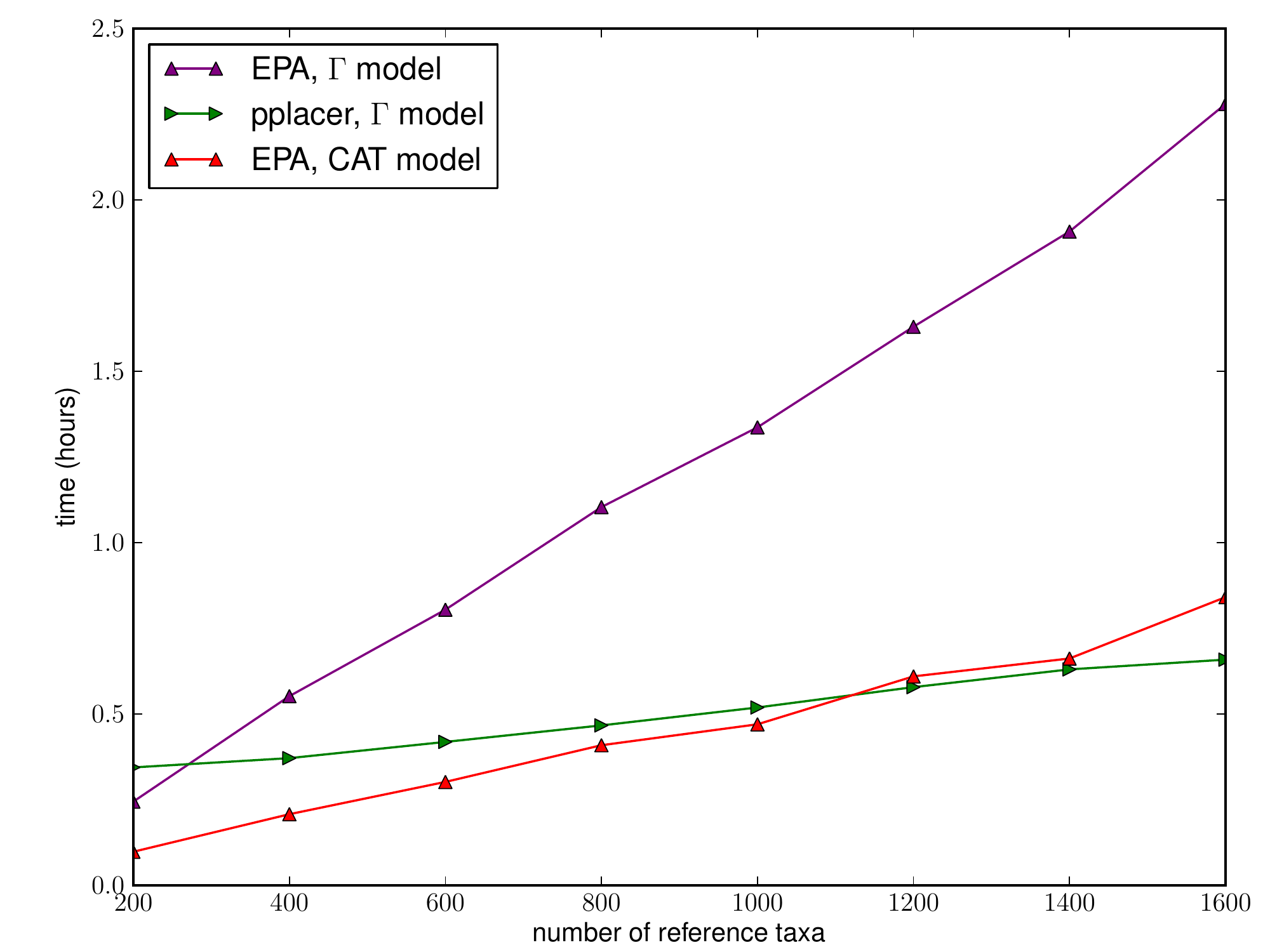}}
\newcommand{\LEGEPA}{
Time to place 10,000 16s rRNA reads of median length 198nt onto a reference phylogenetic tree, with a 1287 nt reference alignment. 
``$\Gamma$ model'' refers to a four-category gamma model of rate heterogeneity \cite{yangRatesAcrossSites94}, and ``CAT'' is an approximation which chooses a single rate for each site \cite{stamatakisCAT06}.
Tests run on an Intel Xeon @ 2.33 Ghz.
}

\newcommand{\tabCOG}{1}
\newcommand{\LEGCOG}{
Error analysis for the COG simulation with the error metric described in the text.
As in Figure~\figmaincog, simulated reads had a normally-distributed length with a mean of 85 amino acids, and a standard deviation of 20.
This table pools the results, and shows mean ($\mu$) and standard deviation ($\sigma$) of the error and the number of reads placed for \pplacer\ run in maximum likelihood (ML) and posterior probability (PP) modes.
For example, the ``ML'' columns in the row labeled 0.4-0.5 shows error statistics for all of the reads in the simulation that had likelihood weight ratio between 0.4 and 0.5: there were 20120 such reads, with error mean and standard deviation of about 2.25 and 2.29, respectively.
This table demonstrates the effectiveness of the confidence scores-- as the confidence scores increase, the error decreases. 
We note that the ML and PP methods have very comparable performance for this length of read, and thus the quickly-calculated ML weight ratio can act as a proxy for the more statistically rigorous posterior probability calculation. 
}

\newcommand{\tabCOGshort}{2}
\newcommand{\LEGCOGshort}{
Similar analysis as Table~\tabCOG, but with a normally-distributed length with a mean of 30 amino acids, and a standard deviation of 7.
In this case, the posterior probability calculation shows slightly superior ability to distinguish between accurate and inaccurate placements than the likelihood weight ratio.
}

\newcommand{\sfigntaxa}{S1}
\newcommand{\sfigalnlen}{S2}

\newenvironment{foldstuff}{}{}

\newboolean{publ}

\newenvironment{bmcformat}{\baselineskip20pt\setboolean{publ}{false}}{\baselineskip20pt}

\begin{document}
\begin{bmcformat}
\begin{foldstuff}
\title{\pplacer: linear time maximum-likelihood and Bayesian phylogenetic placement of sequences onto a fixed reference tree}



\author{Frederick A Matsen\correspondingauthor$^{1}$%
       \email{Frederick A Matsen\correspondingauthor - ematsen@gmail.com}%
      \and
      Robin B Kodner$^2{}^,{}^3$%
         \email{Robin B Kodner - rkodner@u.washington.edu}%
       \ and
         E Virginia Armbrust$^2$%
         \email{E Virginia Armbrust - armbrust@u.washington.edu}%
      }

\address{%
    \iid(1)Departments of Integrative Biology, Mathematics, and Statistics, University of California, Berkeley, Berkeley, CA, USA\\
    \iid(2)School of Oceanography, University of Washington, Seattle, Washington, USA\\
    \iid(3)Friday Harbor Laboratories, University of Washington, Friday Harbor, Washington, USA
}%

\maketitle
\end{foldstuff}

\begin{abstract}
  \paragraph*{Background:} 
  Likelihood-based phylogenetic inference is generally considered to be the most reliable classification method for unknown sequences. 
  However, traditional likelihood-based phylogenetic methods cannot be applied to large volumes of short reads from next-generation sequencing due to computational complexity issues and lack of phylogenetic signal.
  ``Phylogenetic placement,'' where a reference tree is fixed and the unknown query sequences are placed onto the tree via a reference alignment, is a way to bring the inferential power offered by likelihood-based approaches to large data sets.
      
  \paragraph*{Results:} 
  This paper introduces \pplacer, a software package for phylogenetic placement and subsequent visualization.
  The algorithm can place twenty thousand short reads on a reference tree of one thousand taxa per hour per processor, has essentially linear time and memory complexity in the number of reference taxa, and is easy to run in parallel.
  \Pplacer\ features calculation of the posterior probability of a placement on an edge, which is a statistically rigorous way of quantifying uncertainty on an edge-by-edge basis.
  It also can inform the user of the positional uncertainty for query sequences by calculating expected distance between placement locations, which is crucial in the estimation of uncertainty with a well-sampled reference tree.
  The software provides visualizations using branch thickness and color to represent number of placements and their uncertainty.
  A simulation study using reads generated from 631 COG alignments shows a high level of accuracy for phylogenetic placement over a wide range of alignment diversity, and the power of edge uncertainty estimates to measure placement confidence.

  \paragraph*{Conclusions:} 
  \Pplacer\ enables efficient phylogenetic placement and subsequent visualization, making likelihood-based phylogenetics methodology practical for large collections of reads; it is freely available as source code, binaries, and a web service.

\end{abstract}

\ifthenelse{\boolean{publ}}{\begin{multicols}{2}}{}

\section*{Background}

High-throughput pyrosequencing technologies have enabled the widespread use of metagenomics and metatranscriptomics in a variety of fields\cite{margulies2005gsm}.  
This technology has revolutionized the possibilities for unbiased surveys of environmental microbial diversity, ranging from the human gut to the open ocean\cite{culley2006mac, gill2006mah, venter2004egs, tringe2005mds, martin2006mat, warnecke2007maf, baker2003mca}.  
The trade off for high throughput sequencing is that the resulting sequence reads can be short and come without information on organismal origin or read location within a genome.

The most common way of analyzing a metagenomic data set is to use BLAST \cite{altschulEaBLAST90} to assign a taxonomic name to each query sequence based on ``reference'' data of known origin.
This strategy has its problems: when a query sequence is only distantly related to sequences in the database, BLAST can either err substantially by forcing a query into an alignment with a known sequence, or return an uninformatively broad collection of alignments.
Furthermore, similarity statistics such as BLAST $E$-values can be difficult to interpret because they are dependent on fragment length and database size.
Therefore it can be difficult to know if a given taxonomic assignment is correct unless a very clear ``hit'' is found.

Numerous tools have appeared that assign taxonomic information to query sequences, overcoming the shortcomings of BLAST.
For example, MEGAN (MEtaGenome ANalyzer) \cite{husonEaMEGAN07} implements a common-ancestor algorithm on the NCBI taxonomy using BLAST scores.
PhyloPythia \cite{mchardy2007apc}, TACOA \cite{diazEaTACOA09}, and Phymm \cite{bradySalzbergPhymm09} use composition based methods to assign taxonomic information to metagenomic sequences.
Recent tools can work with reads as short as 100bp.

Phylogeny offers an alternative and complementary means of understanding the evolutionary origin of query sequences.
The presence of a query sequence on a certain branch of a tree gives precise information about the evolutionary relationship of that sequence to other sequences in the tree.
For example, a query sequence placed deep in the tree can indicate \emph{how} the query is distantly related to the other sequences in the tree, whereas the corresponding taxonomic name would simply indicate membership in a large taxonomic group.
On the other hand, taxonomic names are key to obtaining functional information about organisms, and the most robust and comprehensive means of understanding the composition of unknown sequences will derive both from taxonomic and phylogenetic sources.

Likelihood-based phylogenetics, with over 30 years of theoretical and practical development, is a sophisticated tool for the evolutionary analysis of sequence data.
It has well-developed statistical foundations for inference \cite{allmanRhodesCovarionMixture06, allmanRhodesGTRI08}, tests for uncertainty estimation \cite{shimodairaHasegawaSHTest99}, and sophisticated evolutionary models \cite{yangRatesAcrossSites94, leGascuelMatrix08}.
In contrast to distance-based methods, likelihood-based methods can use both low and high variation regions of an alignment to provide resolution at different levels of a phylogenetic tree \cite{felsensteinBook04}.

Traditional likelihood-based phylogenetics approaches are not always appropriate for analyzing the data from metagenomic and metatranscriptomic studies.
The first challenge is that of complexity: the maximum likelihood phylogenetics problem is NP-hard \cite{chorTullerMLHard06, rochMLHard06} and thus maximum likelihood trees cannot be found in a practical amount of time with many taxa.
A remarkable amount of progress has been made in approximate acceleration heuristics \cite{guindonGascuelPHYML03,stamatakisEaRAxML06,zwicklGarli06, priceEaFastTree210}, but accurate maximum likelihood inference for hundreds of thousands of taxa remains out of reach.

Second, accurate phylogenetic inference is not possible with fixed length sequences in the limit of a large number of taxa.
This can be seen via theory \cite{steelSzekeleyInvertingRandomII02}, where lower bounds on sequence length can be derived as an increasing function of the number of taxa.
It is clear from simulation \cite{moretEaSequenceLengthRequirements02}, where one can directly observe the growth of needed sequence length.
Such problems can also be observed in real data where insufficient sequence length for a large number of taxa is manifested as a large collection of trees similar in terms of likelihood \cite{bergerStamatakisEPA09}; statistical tools can aid in the diagnosis of such situations \cite{shimodairaHasegawaSHTest99}.

The lack of signal problem is especially pronounced when using contemporary sequencing methods that produce a large number of short reads.
Some methodologies, such as 454 \cite{454}, will soon be producing sequence in the 600-800 bp range, which is sufficient for classical phylogenetic inference on a moderate number of taxa.
However, there is considerable interest in using massively parallel methodologies such as SOLiD and Illumina which produce hundreds of millions of short reads at low cost \cite{mardisNextGenSequencing08}.
Signal problems are further exacerbated by shotgun sequencing methodology where the sequenced position is randomly distributed over a given gene.
Applying classical maximum-likelihood phylogeny to a single alignment of shotgun reads together with full-length reference sequences can lead to artifactual grouping of short reads based on the read position in the alignment; such grouping is not a surprise given that non-sequenced regions are treated as missing data (see, e.g. \cite{felsensteinBook04,lemmon2009effect}). 

A third problem is deriving meaningful information from large trees.
Although significant progress has been made in visualizing trees with thousands of taxa \cite{archaeopteryx, dendroscope}, understanding the similarities and differences between such trees is inherently difficult.
In a setting with lots of samples, constructing one tree per sample requires comparing trees with disjoint sets of taxa; such comparisons can only be done in terms of tree shape \cite{mooersHeardTreeShape97}.
Alternatively, phylogenetic trees can be constructed on pairs of environments at a time, then comparison software such as UniFrac \cite{lozuponeKnightUniFrac05} can be used to derive distances between them, but the lack of a unifying phylogenetic framework hampers the analysis of a large collection of samples.

``Phylogenetic placement'' has emerged in the last several years as an alternative way to gain an evolutionary understanding of sequence data from a large collection of taxa.
The input of a phylogenetic placement algorithm consists of a reference tree, a reference alignment, and a collection of query sequences.
The result of a phylogenetic placement algorithm is a collection of assignments of query sequences to the tree, one assignment for each query.
Phylogenetic placement is a simplified version of phylogenetic tree reconstruction by sequential insertion \cite{klugeFarris69, felsensteinML81}.
It has been gaining in popularity, with recent implementations in 2008 \cite{monierEaLargeViruses08,vonMeringEaQuantitative08}, with more efficient implementations in this paper and by Berger and Stamatakis \cite{bergerStamatakisEPA09}. 
A recent HIV subtype classification scheme \cite{pondEaSCUEAL09} is also a type of phylogenetic placement algorithm that allows the potential for recombination in query sequences.

Phylogenetic placement sidesteps many of the problems associated with applying traditional phylogenetics algorithms to large, environmentally-derived sequence data.
Computation is significantly simplified, resulting in algorithms that can place thousands to tens of thousands of query sequences per hour per processor into a reference tree on several hundred taxa.
Because computation is performed on each query sequence individually, the calculation can be readily parallelized.
The relationships between the query sequences are not investigated, reducing from an exponential to a linear number of phylogenetic hypotheses.
Short and/or non-overlapping query sequences pose less of a problem, as query sequences are compared to the full-length reference sequences.
Visualization of samples and comparison between samples are facilitated by the assumption of a reference tree, that can be drawn in a way which shows the location of reads.

Phylogenetic placement is not a substitute for traditional phylogenetic analysis, but rather an approximate tool when handling a large number of sequences.
Importantly, the addition of a taxon $x$ to a phylogenetic data set on taxa $S$ can lead to re-evaluation of the phylogenetic tree on $S$; this is the essence of the taxon sampling debate \cite{zwicklHillisIncreasedTaxSampGood02} and has recently been the subject of mathematical investigation \cite{cuetoMatsen10}.
This problem can be mitigated by the judicious selection of reference taxa and the use of well-supported phylogenetic trees.
The error resulting from the assumption of a fixed phylogenetic reference tree will be smaller than that when using an assumed taxonomy such as the commonly used NCBI taxonomy, which forms a reference tree of sorts for a number of popular methods currently in use \cite{husonEaMEGAN07, munchEaBarcoding08}.
Phylogenetic placement, in contrast, is done on a gene-by-gene basis and can thus accommodate the variability in the evolutionary history of different genes, which may include gene duplication, horizontal transfer, and loss.

This paper describes \pplacer, software developed to perform phylogenetic placement with linear time and memory complexity in each relevant parameter: number of reference sequences, number of query sequences, and sequence length.
\Pplacer\ was developed to be user-friendly, and its design facilitates integration into metagenomic analysis pipelines.
It has a number of distinctive features. 
First, it is unique among phylogenetic placement software in its ability to evaluate the posterior probability of a placement on an edge, which is a statistically rigorous way of quantifying uncertainty on an edge-by-edge basis.
Second, \pplacer\ enables calculation of the expected distance between placement locations for each query sequence; this development is crucial for uncertainty estimation in regions of the tree consisting of many short branches, where the placement edge may be uncertain although the correct placement region in the tree may be relatively clear. 
Third, \pplacer\ can display both the number of placements on an edge and the uncertainty of those placements on a single tree (Figure~\figapplicationEDPL).
Such visualizations can be used to understand if placement uncertainty is a significant problem for downstream analysis and to identify problematic parts of the tree.
Fourth, the \pplacer\ software package includes utilities to ease large scale analysis and sorting of the query alignment based on placement location.
These programs are available in GPLv3-licensed code and binary form (\hbox{http://matsen.fhcrc.org/pplacer/}), which also includes a web portal for running \pplacer\ and for visualizing placement results.

To validate \pplacer's phylogenetic placement algorithm we implemented a framework that simulates reads from real alignments and tests \pplacer's ability to place the read in the correct location.
As described below, a primary focus of this effort is a simulation study of \ncogs\ COG alignments, where 10 reads were simulated from each taxon of each alignment, placed on their respective trees, and evaluated for accuracy.
These tests confirm both that \pplacer\ places reads accurately and that the posterior probability and the likelihood weight ratio (described below) both do a good job of indicating whether a placement can be trusted or not.
We also use these simulations to understand how the distance to sister taxon impacts placement accuracy.

\arxiv{
\begin{figure}[]
  \begin{center}
    \INCapplicationEDPL
  \end{center}
  \caption{\LEGapplicationEDPL}
\end{figure}
}

\section*{Results}
\subsection*{Overview of phylogenetic placement using \pplacer}

\Pplacer\ places query sequences in a fixed reference phylogeny according to phylogenetic posterior probability and maximum likelihood criteria.
In Bayesian mode, \pplacer\ evaluates the posterior probability of a fragment placement on an edge conditioned on the reference tree topology and branch lengths.
The posterior probability has a clear statistical interpretation as the probability that the fragment is correctly placed on that edge, assuming the reference tree, the alignment, and the priors on pendant branch length.
Because the reference tree is fixed, direct numerical quadrature over the likelihood function can be performed to obtain the posterior probability rather than relying on Markov chain Monte-Carlo procedures as is typically done in phylogenetics \cite{beast, mrBayes}. 
In maximum likelihood (ML) mode, \pplacer\ evaluates the ``likelihood weight ratio,'' \cite{vonMeringEaQuantitative08} i.e. the ML likelihood values across all placement locations normalized to sum to one.

Because the reference tree is fixed with respect to topology and branch length, only two tree traversals are needed to pre-compute all of the information needed from the reference tree. 
From there all likelihood computation is performed on a collection of three-taxon trees, the number of which is linear in the number of reference taxa.
Therefore the fragment placement component of our algorithm has linear ($O(n)$) time and space complexity in the number of taxa $n$ in the reference tree (Figures~\figspeed\ and \figmem).

\arxiv{
\begin{figure}[h]
  \begin{center}
    \INCspeed
  \end{center}
  \caption{\LEGspeed}
\end{figure}

\begin{figure}[h]
  \begin{center}
    \INCmem
  \end{center}
  \caption{\LEGmem}
\end{figure}
}

The \pplacer\ binary is stand-alone; a single command specifying the reference tree, the reference alignment, a reference statistics file, and the aligned reads suffices to run the core \pplacer\ analysis.
\Pplacer\ does not optimize sequence mutation model parameters, and instead obtains those values from PHYML \cite{guindonGascuelPHYML03} or RAxML \cite{stamatakisEaRAxML06} statistics output files.
When analyzing protein sequences the user can choose between the LG \cite{leGascuelMatrix08} or WAG \cite{whelanGoldmanWAG01} models, and nucleotide likelihoods are computed via the general time reversible (GTR) model.
Rate variation among sites is accomodated by the discrete $\Gamma$ model \cite{yangRatesAcrossSites94}.
For posterior probability calculation, the user can choose between exponential or uniform pendant branch length priors.
Each \pplacer\ run creates a \texttt{.place} file that describes the various placements and their confidence scores; analysis can be done directly on this file, or the user can run it through \placeviz, our tool to visualize the fragment placements. 
The \pplacer\ code is written in the functional/imperative language \texttt{ocaml} \cite{ocaml} using routines from the GNU scientific library (GSL) \cite{gsl}.

To accelerate placements, \pplacer\ implements a two-stage search algorithm for query sequences, where a quick first evaluation of the tree is followed by a more detailed search in high-scoring parts of the tree.
The more detailed second search is directed by \pplacer's ``baseball'' heuristics, which limit the full search in a way that adapts to the difficulty of the optimization problem (described in detail in ``Methods'').
The balance between speed and accuracy depends on two parameters, which can be appropriately chosen for the problem at hand via \pplacer's ``fantasy baseball'' mode.
This feature places a subset of the query sequences and reports the accuracy of the parameter combinations within specified ranges, as well as information concerning runtime for those parameter combinations.
The user can then apply these parameter choices for an optimized run of their data.

\subsection*{Quantifying uncertainty in placement location}
 
\Pplacer\ calculates edge uncertainty via posterior probability and the likelihood weight ratio.
These methods quantify uncertainty on an edge-by-edge basis by comparing the best placement locations on each edge.
Such quantities form the basis of an understanding of placement uncertainty.

The Expected Distance between Placement Locations (EDPL) is used to overcome difficulties in distinguishing between local and global uncertainty, which is a complication of relying on confidence scores determined on an edge-by-edge basis.
This quantity is computed as follows for a given query sequence.
\Pplacer\ first determines the top-scoring collection of edges; the optimal placement on each edge is assigned a probability defining confidence, which is the likelihood weight ratio (in ML mode) or the posterior probability (in Bayesian mode). 
The EDPL uncertainty is the weighted-average distance between those placements (Figure~\figEDPL), i.e. the sum of the distances between the optimal placements weighted by their probability (\ref{eq:EDPL}).
The EDPL thus uses distances on the tree to distinguish between cases where nearby edges appear equally good, versus cases when a given query sequence does not have a clear position in the tree.
These measures of uncertainty can then be viewed with \placeviz\ as described below.

\arxiv{
\begin{figure}[]
  \begin{center}
    \INCEDPL
  \end{center}
  \caption{\LEGEDPL}
\end{figure}
}

\subsection*{Visualizing placements using \placeviz\ and placement management using \placeutil}

Our package includes tools to facilitate placement visualization and management: \placeviz\ and \placeutil.
\Placeviz\ converts the placement files generated by \pplacer\ into tree formats that are viewable by external viewers.
The richest visualizations make use of the phyloXML format \cite{hanZmasekPhyloXML09}, which can be viewed using the freely available Archaeopteryx \cite{archaeopteryx} Java software.
Less information-dense visualizations are also available in the standard ``Newick'' format \cite{felsensteinBook04}.

As shown in Figure~\figapplicationEDPL, \placeviz\ extends previous work on visualizations \cite{vonMeringEaQuantitative08}, representing placement density (branch thickness) and uncertainty (color) on a single tree.
Specifically, it draws the reference tree such that the thickness of the branch is a linear function of the number of placements (this linear function has a non-zero $y$-intercept so that the whole tree is visible); the weighted average EDPL uncertainty for the placements on the tree is expressed as a color gradient from the usual branch length color (white or black by choice) to red, with 100\% red representing a user-defined uncertainty maximum.
Yellow is used to denote edges whose average EDPL uncertainty is above the given maximum level.

\Placeviz\ also offers other visualization options, such as individually placing the query sequences on the tree, which is useful for a small number of placements.
It also can sort query sequences by their best scoring edge into a \texttt{.loc.fasta} file; inspection can reveal if any specific features of the query sequences lead to placement on one edge or another.
This sorting can also group query sequences as potentially coming from similar organisms, even if those query sequences do not overlap.

\Placeutil\ is a utility for combining, splitting apart, and filtering placements, which can be useful when doing large scale analysis.
For example, when a collection of query sequences are split apart to run in parallel, their placements can be brought back together using \placeutil, while checking that they were run using the same reference tree and model parameters.
Conversely, if a number of samples were run together, they can be split apart again using regular expressions on their names.
Placements can also be separated by likelihood weight ratio, posterior probability, and EDPL.

\subsection*{A \pplacer\ application: \psbA\ in the Global Ocean Sampling (GOS) database}
To demonstrate the use of \pplacer for a metagenomic study, we analyzed the \psbA\ and \psbD\ gene for the D1 and D2 subunits of photosystem II in cyanobacterial and eukaryotic chloroplasts \cite{zurawski1982nucleotide} from the Global Ocean Sampling (GOS) dataset \cite{venter2004egs}.
The GOS database is the largest publicly available metagenomic database, and has been the subject of numerous studies.
We choose the \psbA\ and \psbD\ genes because they are well defined, are found across domains, and can be used to differentiate cyanobacteria from eukaryotic phototrophs in a data set assuming sequence reads are accurately identifed \cite{zeidner2003molecular}.
In addition, it has been shown in a number of studies that cyanophage virus genomes contain both \psbA\ and \psbD\ sequences \cite{sullivan2006prevalence, millard2004genetic, lindell2007genome, chenard2008phylogenetic}, and that viruses are the source of a substantial number \psbA\ and \psbD\ sequences in GOS \cite{williamson2008sorcerer, sharon2007viral}.
BLAST results on the GOS database with TBLASTN, BLASTN, or BLASTP using eukaryotic query sequences were similar to those retrieved when using cyanobacterial query sequences, making BLAST-based taxonomic identification difficult, even at a high taxonomic level.
The use of \pplacer\ on the closely related \psbA\ and \psbD\ genes demonstrates phylogenetic placement on closely related paralogs.

To identify \psbA\ and \psbD\ genes in the GOS dataset, we performed a HMMER \cite{eddyProfileHMMs98} search of the GOS dataset using a 836 nucleotide reference alignment containing 270 reference sequences of cyanobacteria, eukaryotic plastids, and virus.
The reference alignment included all possible reference sequences for \psbA\ and \psbD\ from published genomes, which is important for confident phylogenetic identification of new clades or strains.
A total of 8535 metagenomic sequences were recruited by HMMER with an E-value cut off of $10^{-5}$; these were then placed on the reference tree using \pplacer\ (Figures~\figapplicationEDPL\ and \figapplication).
The expanded region of the trees shown in the figures highlights the \pro\ clade, known to be one of the most abundant phototrophs in the global ocean.
There are many sequences placed sister to the sequenced representatives but also many sequences placed at internal nodes, that could represent some as yet unsequenced strain of these cyanobacteria.

\arxiv{
\begin{figure}[]
  \begin{center}
    \INCapplication
  \end{center}
  \caption{\LEGapplication}
\end{figure}
}

\subsection*{Simulation}

Simulation experiments were conducted to verify overall accuracy and to determine the relationship between confidence scores and accuracy.
The simulation removes one taxon at a time from a given reference tree, simulates fragments from that taxon, then evaluates how accurately the placement method assigns the simulated fragments to their original position.
In order to evaluate the accuracy of the placements, a simple topological distance metric is used.
We have not simulated homopolymer-type errors in the alignments, because such errors should be treated by a pre-processing step and thus are not the domain of a phylogenetic placement algorithm.
Furthermore, the emergence of more accurate very high throughput sequencing technology \cite{mardisNextGenSequencing08} re-focuses our attention on the question of speed rather than error problems.
Further details are given in the ``Methods'' section.

A broad simulation analysis of \pplacer\ performance was done using \ncogs\ COG \cite{tatusov2000cdt} alignments.
The COG alignments had between 19 and 436 taxa, with a median of 41; they were between 200 and 2050 amino acids in length, with a median of 391 (supplemental Figures~\sfigntaxa\ and \sfigalnlen).
Reference phylogenetic trees were built based on the full-length gene sequences for each of these genes using PHYML \cite{guindonGascuelPHYML03} and the LG \cite{leGascuelMatrix08} protein substitution model.
Each taxon from each gene alignment was eliminated one at a time from the reference set as described in ``Methods''; ten reads were simulated from each, leading to a total of 334,670 simulated reads, which were aligned to a hidden Markov model of the reference alignment. 
As is commonly done when analyzing a metagenome, the reads were filtered by their HMMER E-value (in this case $10^{-5}$).
Two normal read length distributions were used: a ``long'' read simulation with amino acid sequence length of mean 85 and standard deviation of 20, and a ``short'' read simulation with mean 30 and standard deviation of 7.
After the HMMER step, the ``long'' read simulation placed a total of 285,621 reads, and the ``short'' one placed a total of 148,969 reads on their respective phylogenetic trees. 

The best resulting maximum likelihood placement edge was compared to the placement with the highest posterior probability to determine how well the confidence scores reflect the difference between accurate and inaccurate placements (Tables~\tabCOG\ and \tabCOGshort).
Both methods provide similar results, implying that the likelihood weight ratio appears to be a reasonable proxy for the more statistically rigorous posterior probability calculation, although posterior probability does a slightly better job of distinguishing between accurate and inaccurate placements for the short reads.
Overall, accuracy is high as there is a strong correlation between likelihood weight ratio, posterior probability, and accuracy.
Many of the placements were placed with high confidence score and high accuracy in large and small trees (Figure~\figmaincog).
Reads from more closely related taxa are easier to accurately place than more distantly related taxa (Figure~\figsistercog), although good placement is achieved even when sequences are only distantly related to the sequences in the reference tree. 

\arxiv{
\begin{figure}[]
  \begin{center}
    \INCmaincog
  \end{center}
  \caption{\LEGmaincog}
\end{figure}

\begin{figure}[]
  \begin{center}
    \INCsistercog
  \end{center}
  \caption{\LEGsistercog}
\end{figure}
}

\section*{Discussion}

Likelihood-based phylogeny is a well developed way to establish the evolutionary relationships between sequences.
Phylogenetic placement is a simplified version of likelihood-based phylogenetic inference that enables rapid placement of numerous short query sequences and which sidesteps some of the problems inherent in applying phylogenetics to hundreds of thousands or millions of taxa.
Phylogenetic placement is by no means a replacement for classical phylogenetic inference, which should be applied when query sequences are full length and moderate in number.

Phylogenetic placement software sits in a category distinct from taxonomic identification software such as MEGAN \cite{husonEaMEGAN07} or Phymm \cite{bradySalzbergPhymm09}.
First, phylogenetic placement software does not assign names to query sequences, and instead returns an assignment of the query sequences to edges of a phylogenetic tree.
Second, phylogenetic placement is designed for fine-scale analysis of query sequences to provide detailed comparative and evolutionary information at the single gene level.
This poses no problems when looking at a single marker gene such as such as 16S, but some scripting and automation is necessary when there are many genes of interest. 
These challenges are somewhat mitigated through program design and pipeline scripts \cite{starkEaMLTreeMap10}, but phylogenetic placement methods may always require more work than general purpose taxonomic classification software.

Phylogenetic placement is also different than packages which construct a phylogenetic tree \emph{de novo} in order to infer taxonomic identity by clade membership.
Such packages, such as CARMA \cite{krause2008pcs} and SAP \cite{munchEaBarcoding08, munchEaSAP08}, combine sequence search, alignment, and phylogeny into a complete pipeline to provide taxonomic information for an unknown query sequence.
Because different query sequences will have different sets of reference taxa, these methods are not phylogenetic placement algorithms as described above.
Also, because they are performing a full phylogenetic tree construction, they either use distance-based methods for faster results \cite{krause2008pcs,munchEaBarcoding08} or are many orders of magnitude slower than phylogenetic placement methods \cite{munchEaSAP08}.

\Pplacer\ is not the only software to perform likelihood-based phylogenetic placement.
The first pair of software implementations were the ``phylomapping'' method of \cite{monierEaLargeViruses08}, and the first version of the ``MLTreeMap'' method of \cite{vonMeringEaQuantitative08}. 
Both methods use a topologically fixed reference tree, and are wrappers around existing phylogenetic implementations: ProtML \cite{felsensteinPhylip04} for phylomapping, and TREE-PUZZLE \cite{schmidtEaTREEPUZZLE02} for MLTreeMap.
Neither project has resulted in software that is freely available for download (MLTreeMap is available as a web service, but as it is tied to a core set of bacterial genes it is not useful for scientists examining other genes or domains).
Also, by using a general-purpose phylogenetic computing engine, they miss on opportunities to optimize on computation and the resulting algorithm is not linear in the number of reference taxa.
Both methods equip placement with a statistically justifiable but non-traditional confidence score: phylomapping adapts the RELL bootstrap \cite{kishinoEaRELL90} to their setting, and MLTreeMap uses the ``expected maximum likelihood weight ratio,'' which has been discussed in \cite{strimmerRambautConfidence02}.
AMPHORA also uses a hybrid parsimony and neighbor-joining strategy to place query sequences in a fixed reference tree \cite{wu2008sfa}.

The only other software at present that performs likelihood-based phylogenetic placement at speeds comparable of \pplacer\ is the independently-developed ``evolutionary placement algorithm'' (EPA) \cite{bergerStamatakisEPA09} available as an option to RAxML \cite{stamatakisEaRAxML06}.
In a comparison designed jointly by ourselves and the authors of \cite{bergerStamatakisEPA09}, \pplacer\ and the EPA showed comparable speed (Figure~\figEPA).
\Pplacer\ and the EPA both cache likelihood information on the tree to accelerate placement, and both use two-stage algorithms to quickly place many sequences.
The two packages use different acceleration heuristics, but only \pplacer\ offers guidance on parameter choices to use for those heuristics via its  ``fantasy baseball'' feature as described above.
The EPA allows for one parameter more flexibility than \pplacer\ for branch length optimization, and can perform placement on partitioned datasets and inference on binary, RNA secondary structure, and multi-state data.
The EPA offers single-process parallelization (note both the EPA and \pplacer\ can easily be run in parallel as multiple processes).
The EPA comes without a visualization tool such as \placeviz, although it can be run within the new MLTreeMap suite of Perl scripts for visualization \cite{starkEaMLTreeMap10}.

\arxiv{
\begin{figure}[]
  \begin{center}
    \INCEPA
  \end{center}
  \caption{\LEGEPA}
\end{figure}
}

\section*{Conclusions}

\Pplacer\ enables efficient maximum likelihood and posterior probability phylogenetic placement of reads, making likelihood-based phylogenetics methodology practical for large-scale metagenomic or 16S survey data.
\Pplacer\ can be used whenever a reference alignment and phylogenetic tree is available, and is designed for ease of use for both single-run and pipelined applications.
``Baseball'' heuristics adapt to the difficulty of the phylogenetic placement problem at hand, and come with features which guide the user to an appropriate set of parameter choices.
The EDPL metric helps users decide if edge uncertainty is a substantial problem for downstream analysis.
\Pplacer\ offers tightly integrated yet flexible visualization tools which can be used to view both the placements and their uncertainty on a single tree.
Large-scale simulations confirmed the accuracy of the \pplacer\ results and the descriptive ability of the confidence scores.
\Pplacer\ is freely available, comes with a complete manual and tutorials, and can be used via a web service.

\Pplacer\ forms the core of a body of work we are developing to facilitate and extend the utility of phylogenetic placement methodology.
The next step will be to release software implementing a metric similar to UniFrac \cite{lozuponeKnightUniFrac05} which allows for statistical comparison and visualization of differences between samples.
In collaboration with another group, we have also implemented a preliminary version of software which automates the selection of appropriate reference sequences, as well as the assignment of taxonomic names based on phylogenetic placements.
Further releases of \pplacer\ will also implement low-level optimizations for increased speed.
 
\section*{Methods}
\subsection*{\Pplacer\ algorithmic internals}
Here we survey \pplacer\ algorithmic developments. 
The code implementing these algorithms is freely available on the github code repository \cite{pplacer_github}.
The basic development that permits linear time and space scaling in the size of the reference tree is that of pre-calculation of likelihood vectors at either end of each edge of the reference tree; this development is shared by the EPA and SCUEAL \cite{pondEaSCUEAL09} and the original idea goes back much earlier.
Using these cached likelihood vectors, a naive algorithm might insert the query sequence into each edge of the tree and perform full branch length optimization using the cached likelihood vectors.
However, a substantial speed improvement can be gained by performing a two-stage algorithm, where the first stage does a quick initial evaluation to find a good set of locations, and the second stage does a more detailed evaluation of the results from the first stage.

\Pplacer's ``baseball'' heuristics limit the full search on the tree in a way that adapts to the difficulty of the optimization problem.
The first stage is enabled by calculating likelihood vectors for the center of each edge; these vectors can be used to quickly sort the edges in approximate order of fit for a given query sequence.
This edge ordering will be called the ``batting order.''
The edges are evaluated in the batting order with full branch length optimization, stopping as follows.
Start with the edge that looks best from the initial evaluation; let $L$ be the log likelihood of the branch-length-optimized ML attachment to that edge.
Fix some positive number $D$, called the ``strike box.''
We proceed down the list in order until we encounter the first placement that has log likelihood less than $L - D$, which is called a ``strike.''
Continue, allowing some number of strikes, until we stop doing detailed evaluation of what are most likely rather poor parts of the tree.
An option restricts the total number of ``pitches,'' i.e. full branch length optimizations.

The baseball heuristics allow the algorithm to adapt to the likelihood surface present in the tree; its behavior is controlled by parameters that can be chosen using \pplacer's ``fantasy baseball'' feature.
This option allows automated testing of various parameter combinations for the baseball heuristics.
Namely, it evaluates a large fixed number of placements, and records what the results would have been if various settings for the number of allowed strikes and the strike box were chosen.
It records both the number of full evaluations that were done (which is essentially linearly proportional to the run time) and statistics that record if the optimal placement would have been found with those settings, and how good the best found with those settings is compared to the optimal placement.

Placement speed is also accelerated by using information gained about the placement of a given query sequence to aid in placement of closely related query sequences.
Before placement begins, pairwise sequence comparisons are done, first in terms of number of mismatches and second in terms of number of matches to gaps.
Specifically, each sequence $s_i$ is compared to previous sequences in order; the sequence $s_j$ which is most closely related to $s_i$ with $j < i$ is found and assigned as $s_i$'s ``friend.''
If no sequence is found which is above a certain threshold of similarity then no friend is assigned.
If $s_i$ and $s_j$ are identical, then $s_j$'s placement is used for $s_i$. 
If they are similar but not identical, the branch lengths for $s_j$ are used as starting values for the branch length optimization of $s_i$.
This scheme is not a heuristic, but rather an exact way to accelerate the optimization process.

\Pplacer's speed is also linearly proportional to the lengths of the query sequences, which is enabled because the reference tree is fixed with respect to topology and branch length.
Specifically, as described below, likelihood computations are performed such that the sites without a known state (gaps or missing sites) cancel out of the computation of likelihood weight or posterior probability.
These sites are masked out of \pplacer's computation and thus do not compute to runtime.

Because of the extensive memory caching to accelerate placement, \pplacer\ consumes a nontrivial amount of memory.
The fixed contributions to memory use break down as follows: 
a factor of two for quick and full evaluation of placements,
two nodes on each edge,
four rate variation categories,
four bytes per double precision floating point number,
and four (nucleotide) or 20 (amino acid) states.
To get a lower bound for total memory use, multiply this number, which is 128 bytes (nucleotide) or 640 bytes (amino acid), with two times the number of reference sequences minus three (the number of edges), times the number of columns in the reference alignment.
Other data structures add on top of that (Figure~\figmem).

\subsection*{Likelihood weight ratio, posterior probability, and EDPL}
Posterior probability is calculated by first integrating out the possible attachment locations and branch lengths against a prior distribution of pendant branch lengths.
Let $\ell_i$ denote an edge of the reference tree, $A_i$ the length of that edge, $a$ the attachment location along $\ell_i$, $b$ the pendant branch length, $\Lcal$ the phylogenetic likelihood function (e.g. equation 16.9 of \cite{felsensteinBook04}), $D$ the alignment, $\Tref$ the reference phylogenetic tree, and $P$ the prior probability of a pendant branch length.
We obtain the Bayes marginal likelihood by direct two-dimensional numerical integration:
\begin{equation}
  \Lbayes(\ell_i | \Tref, D) = A_i^{-1} \int_0^\infty \int_0^{A_i} \Lcal(D | \Tref, \ell_i, a, b) P(b) da \, db
  \label{eq:bayesMarginal}
\end{equation}
The posterior probability can then be obtained by taking a ratio of these marginal likelihoods:
\begin{equation}
  \PP(\ell_i | \Tref, D) = \frac{\Lbayes(D | \Tref, \ell_i)}{\sum_j \Lbayes(D | \Tref, \ell_j)}
  \label{eq:bayesPP}
\end{equation}
The likelihood weight distribution is defined as the corresponding ratio with marginal likelihood replaced by the ML likelihood:
\begin{equation}
  \PP(\ell_i | \Tref, D) =  \frac{\Lml(D | \Tref, \ell_i)}{\sum_j \Lml(D | \Tref, \ell_j)}
  \label{eq:likeWeightDistr}
\end{equation}
The expected (under bootstrap replicates) likelihood weight distribution is the confidence score used in \cite{vonMeringEaQuantitative08}. 
Some justification for using the likelihood weight distribution is given in \cite{strimmerRambautConfidence02}.

The expected distance between placement locations (EDPL) is a simple summation given probabilities from likelihood weight distributions or posterior probabilities.
Let $p_i = \PP(\ell_i | \Tref, D)$ from either (\ref{eq:bayesPP}) or (\ref{eq:likeWeightDistr}), let $d_{ij}$ denote the tree distance between the optimal attachment positions on edges $\ell_i$ and $\ell_j$, and let $L$ denote the total tree length.
Then the EDPL is simply
\begin{equation}
  \sum_{ij} p_i p_j d_{ij} / L
\label{eq:EDPL}
\end{equation}
An extension of these ideas would be to integrate the marginal likelihoods over the potential attachment positions on the edges of interest; we have not pursued such a calculation.

\subsection*{Simulation design and error metric}
The simulation procedure for a single gene is as follows.
Begin with an alignment $A$ of full-length sequences for the gene of interest, along with a phylogeny $T$ derived from that alignment.
$T$ is assumed to be correct. 
Simulated fragments from a given taxon $X$ are re-placed in the phylogenetic tree, and their location relative to $X$'s original location is determined.
The simulation pipeline repeats the following steps for every taxon $X$ in the alignment $A$.
\begin{enumerate}
  \item remove $X$ from the reference alignment, making an alignment $A_X$.
  \item build a profile HMM out of $A_X$.
  \item cut $X$ and its pendant branch out of the tree $T$, suppressing the resultant degree-two internal node. Re-estimate branch lengths using $A_X$, and call the resulting tree $T_X$.
  \item simulate fragments from the unaligned sequence of $X$ by taking sequences of normally-distributed length and uniformly-distributed position.
  \item align these simulated fragments using the profile HMM built from $A_X$. 
  \item place the simulated fragments in $T_X$ with respect to the reference alignment $A_X$.
  \item compare the resulting placements to the location of $X$ in $T$ using our error metric described below.
\end{enumerate}
Note that only branch lengths are re-estimated; if we estimated $T_X$ \emph{de novo} from $A_X$ then we would not be able to compare the placements to the taxon locations in $T$.

In order to evaluate the accuracy of the placements, a simple topological distance metric is used.
To calculate this metric for the placement of a taxon $X$, highlight both the edge of $T_X$ corresponding to the correct placement and the edge of $T_X$ corresponding to the actual placement of the simulated fragment. 
The error metric then is the number of internal nodes between the two highlighted edges.
Thus, if the fragment is placed in the correct position, then error is zero, and if it is placed sister to the correct position, then the error is one, and so on.

\subsection*{Alignments and Reference Trees}
Data for the analysis of speed and memory use was drawn from \cite{turnbaughEaGutMicrobiome08}.
The data came partitioned into two files, the smaller of which was used for the reference set.
Sequences with at least 1200 non-gap characters were selected from the reference set and the sequence order was randomized.
Reference trees were built on the first 200, 400, \dots, 1600 sequences, and the other file was used as the query set.

Alignments for the COG simulation were downloaded from the COG website \cite{tatusov2000cdt}.
The alignments were screened for completeness and taxa with incomplete sequences were removed.  
Alignment ends were trimmed to eliminate excessive gaps on either end.
For the GOS \psbA\ analysis, the - All\_Metagenomic\_Reads and All\_Assembled\_Sequences - were downloaded to a local computer cluster from CAMERA \cite{camera}.  
A \psbA\ and \psbD\ reference alignment was made of eukaryotic plastid sequences using sequences retrieved from Genbank and then included all cyanobacteria with an HMM search of a local copy of microbial refseq (from Genbank); alignment of was done using Geneious alignment \cite{drummond2007gv} and was hand edited.  

\section*{Authors contributions}
FAM and RBK conceived of and developed the project.  
FAM did the coding, scripting, and simulation data analysis.  
RBK made and edited alignments and reference trees.  
FAM, RBK, and EVA analyzed the results and wrote the manuscript.
\section*{Acknowledgements}
  \ifthenelse{\boolean{publ}}{\small}{}
  
  Jonathan Eisen, Steve Evans, John Huelsenbeck and Rasmus Nielsen made helpful suggestions concerning phylogenetics, while Robert Bradley, Ruchira Datta, and Sean Eddy helped with the use of profile HMMs in this context.  
  We thank the Center for Environmental Genomics at the University of Washington, in particular Chris Berthiaume and David Schruth, for computational assistance.
  David Schruth, Adrian Marchetti and Alexandros Stamatakis provided helpful suggestions on the manuscript.
  Simon Berger and Alexandros Stamatakis generously helped with simulation design and running of the EPA algorithm.
  FAM is especially grateful to Andr\'{e}s Var\'{o}n and Ward Wheeler, who made a number of suggestions which greatly improved the \pplacer\ code.
  The following individuals from the \texttt{ocaml} listserv made helpful suggestions:
  Will M. Farr, Mauricio Fernandez, St\'{e}phane Glondu, Jon D. Harrop,  Xavier Leroy, Mike Lin, and Markus Mottl. 
  FAM is supported by the Miller Institute for Basic Research at the University of California, Berkeley.
  RBK is supported by the University of Washington Friday Harbor Laboratories.
  EVA is supported through a Gordon and Betty Moore Foundation Marine Microbiology Investigator award.

{\ifthenelse{\boolean{publ}}{\footnotesize}{\small}
 \bibliographystyle{bmc_article}  
  \bibliography{pplacer} }     


\ifthenelse{\boolean{publ}}{\end{multicols}}{}



\newpage
\bmcSubmit{
\section*{Figures}
\subsection*{Figure~\figapplicationEDPL: Example application, showing uncertainty}
\bmcNoSubmit{\INCapplicationEDPL}

\LEGapplicationEDPL

\bmcNoSubmit{\newpage}

\subsection*{Figure~\figspeed: Linear time dependence on number of reference taxa}
\bmcNoSubmit{\INCspeed}

\LEGspeed

\bmcNoSubmit{\newpage}

\subsection*{Figure~\figmem: \pplacer\ memory requirements}
\bmcNoSubmit{\INCmem}

\LEGmem

\bmcNoSubmit{\newpage}

\subsection*{Figure \figEDPL: measuring uncertainty by the expected distance between placement locations (EDPL)}
\bmcNoSubmit{\INCEDPL}

\LEGEDPL

\bmcNoSubmit{\newpage}

\subsection*{Figure~\figapplication: Example application}
\bmcNoSubmit{\INCapplication}

\LEGapplication

\bmcNoSubmit{\newpage}

\subsection*{Figure~\figmaincog: Simulation with \ncogs\ COG alignments}
\bmcNoSubmit{\INCmaincog}

\LEGmaincog

\bmcNoSubmit{\newpage}

\subsection*{Figure~\figsistercog: Accuracy versus distance to sister taxon: COG simulation}
\bmcNoSubmit{\INCsistercog}

\LEGsistercog

\bmcNoSubmit{\newpage}

\subsection*{Figure~\figEPA: Speed comparison of \pplacer\ and RAxML's EPA algorithm}
\bmcNoSubmit{\INCEPA}

\LEGEPA

\bmcNoSubmit{\newpage}
}


\newpage

\section*{Tables}

\subsection*{Table~\tabCOG\ - Accuracy results for the mean 85 AA COG simulation}

\begin{tabular}{c|cccccc}
range & ML $\mu$ & PP $\mu$ & ML  $\sigma$ & PP  $\sigma$ & ML \# & PP \# \\
\hline
0.00-0.10  &  0.00  &  0.00  &  0.00  &  0.00  &  0       &  0       \\
0.10-0.20  &  3.57  &  3.78  &  3.09  &  3.27  &  4149    &  2312    \\
0.20-0.30  &  2.97  &  3.19  &  3.04  &  3.06  &  15123   &  9018    \\
0.30-0.40  &  2.39  &  2.76  &  3.00  &  3.07  &  22696   &  18373   \\
0.40-0.50  &  2.25  &  2.29  &  3.11  &  2.98  &  20120   &  23022   \\
0.50-0.60  &  2.14  &  2.11  &  3.09  &  3.01  &  17228   &  20090   \\
0.60-0.70  &  1.94  &  1.95  &  3.04  &  2.99  &  14113   &  16223   \\
0.70-0.80  &  1.86  &  1.85  &  3.05  &  3.01  &  13527   &  14879   \\
0.80-0.90  &  1.62  &  1.65  &  2.97  &  2.97  &  14850   &  15747   \\
0.90-1.00  &  0.32  &  0.32  &  1.54  &  1.53  &  163815  &  165957  \\
\end{tabular}

\hspace{0.5cm}

\LEGCOG

\newpage

\subsection*{Table~\tabCOGshort\ - Accuracy results for the mean 30 AA COG simulation}

\begin{tabular}{c|cccccc}
range & ML $\mu$ & PP $\mu$ & ML  $\sigma$ & PP  $\sigma$ & ML \# & PP \# \\
\hline
0.00-0.10  &  0.00  &  0.00  &  0.00  &  0.00  &  0      &  0      \\
0.10-0.20  &  3.67  &  3.94  &  3.23  &  3.31  &  7736   &  3583   \\
0.20-0.30  &  3.24  &  3.48  &  3.26  &  3.23  &  17491  &  14308  \\
0.30-0.40  &  2.64  &  2.98  &  3.23  &  3.26  &  17000  &  17600  \\
0.40-0.50  &  2.51  &  2.46  &  3.30  &  3.11  &  11114  &  14572  \\
0.50-0.60  &  2.27  &  2.27  &  3.26  &  3.10  &  8375   &  9894   \\
0.60-0.70  &  2.11  &  2.03  &  3.14  &  3.08  &  6921   &  7771   \\
0.70-0.80  &  1.83  &  1.76  &  3.06  &  2.98  &  6321   &  6530   \\
0.80-0.90  &  1.51  &  1.44  &  2.92  &  2.83  &  7101   &  6873   \\
0.90-1.00  &  0.22  &  0.20  &  1.22  &  1.17  &  66910  &  67838  \\
\end{tabular}

\hspace{0.5cm}

\LEGCOGshort

\newpage

\end{bmcformat}
\end{document}